\documentclass[reprint, amsmath,amssymb,]{revtex4-2}
\usepackage{graphicx}% Include figure files
\usepackage{dcolumn}% Align table columns on decimal point
\usepackage{bm}% bold math
\usepackage{hyperref}
\usepackage{color}
\usepackage{siunitx}
\usepackage{mathrsfs}

\makeatletter
\newcommand*\bigcdot{\mathpalette\bigcdot@{.7}}
\newcommand*\bigcdot@[2]{\mathbin{\vcenter{\hbox{\scalebox{#2}{$\m@th#1\bullet$}}}}}
\makeatother
\makeatletter
\usepackage[explicit]{titlesec}

\titlespacing*{\section}{0pt}{4.0ex plus .8ex minus 0.5ex}{1.6ex plus .0ex}
\titlespacing*{\subsection}{0pt}{3.5ex plus .0ex minus .0ex}{2.3ex plus .0ex}
\hypersetup{hypertex=true, colorlinks=true, linkcolor=blue, anchorcolor=blue, citecolor=blue}

\begin{document}
	
	\preprint{APS/123-QED}
	
	\title{Mechanical Anisotropy and Multiple Direction-Dependent Dirac States in the Synthesized Ag$_3$C$_2$$_0$ Monolayer}
	
	\author{Zhiheng Ly\textsuperscript{1}, Ningning Jia\textsuperscript{1},  
		Jiangtao Cai\textsuperscript{2}, Jijun Zhao\textsuperscript{3}}
    \email{zhaojj@dlut.edu.cn(J.Zhao)}
    
    \author{Zhifeng Liu\textsuperscript{1}}
	\email{zfliu@imu.edu.cn(Z.Liu)}
	
	\affiliation{%
		$^1$School of Physical Science and Technology, Inner Mongolia
		University, Hohhot 010021, China\\
		$^2$Department of Physics, Shaanxi University of Science and Technology, Xi’an
		710021, China\\
	    $^3$Key Laboratory of Materials Modification by Laser, Ion and Electron Beams
	    (Dalian University of Technology), Ministry of Education, Dalian 116024, China}
    
\date{\today}
	
\begin{abstract}
Recently, a 2D orthorhombic silver-organic framework, Ag$_3$C$_2$$_0$ monolayer, was synthesized by assembling organic molecules linked with multiple aryl-metal bonds. Herein, via first-principles study, we demonstrate that owing to the unique bonding feature, Ag$_3$C$_2$$_0$ monolayer not only exhibits strong mechanical anisotropy, but also possesses various tunable direction-dependent Dirac states. Around the Fermi level blow, the intrinsic Dirac points form two antiparallel quasi type-III nodal lines protected by mirror symmetry, which can further evolve into hybrid nodal loops under tiny strains.  Intriguingly, near the Fermi level above, a special semi-Dirac state can emerge under a critical strain by merging two type-I Dirac cone, which harbors direction-dependent strongly localized fermions, normal massive carries, and ultrafast Dirac fermions at the same time. These findings suggest that the mechanically sensitive Ag$_3$C$_2$$_0$ monolayer is a promising 2D material to realize the interesting Dirac physics and highly anisotropic multiple carries transport.
\end{abstract}
\maketitle
\UseRawInputEncoding

\section{INTRODUCTION}
2D Dirac materials, possessing linear band dispersion near the Fermi level like that of graphene  \cite{2005Novoselov}, have attracted great attention due to many exotic physical phenomena and properties \cite{2015Wang,2022Fan,2014Wehling} (\emph{e.g.}, ballistic transport, high carrier mobility and topological phases) associated with the Dirac states. For versatile applications, the tailorable band dispersions are highly desirable \cite{2010Rusponi}, which determines the tunability of the charge carrier group velocities. Beyond the isotropic 2D Dirac materials, more and more efforts \cite{2010Rusponi,2008Anisotropic,2012Competition,2013dimensional,2015Phagraphene,2015Highly,2018Discovery} 
have been devoted to the 2D systems with anisotropic Dirac band dispersions, such as the graphene superlattices 
\cite{2010Rusponi,2008Anisotropic}, 6,6,12-graphyne \cite{2012Competition}, OPG-Z \cite{2013dimensional}, phagraphene \cite{2015Phagraphene}, S-graphyne \cite{2015Highly} and $\chi$$_3$ borophene \cite{2018Discovery}.

In terms of the anisotropy of band dispersions, Dirac nodal line \cite{2017Line,2017dirac,2017Lu} and semi-Dirac \cite{PhysRevLett.103.016402,PhysRevLett.102.166803,kim2015observation,zhang2017dirac} 
semimetals are particularly outstanding in 2D Dirac materials. For the former, their Dirac points form extended lines in the 2D Brillouin zone (BZ) \cite{2017Lu,PhysRevLett.103.016402,PhysRevLett.102.166803,kim2015observation,zhang2017dirac,feng2017experimental}. Due to the continuity of the Dirac points, band dispersions near one point of them are highly anisotropic, at least being quite different along the tangential and transverse directions, like the case of 3D nodal line semimetals \cite{chang2019realization}. As is known, the linear dispersion along a certain \textit{\textbf{k}} path can usually be classified into three scenarios: (i) the slopes of the two crossing bands are opposite in sign, termed as type-I Dirac dispersion; (ii) the slopes share the same sign, type-II; (iii) the slope of one linear band is equal to zero, while that of the other one is nonzero, type-III. According to these Dirac dispersions, there are three types of nodal lines, namely type-I, -II, and -III nodal lines \cite{he2018type}, in which the band dispersion along the transverse directions of the nodal line are type-I, -II , and -III Dirac dispersions, respectively.

As for the semi-Dirac state \cite{PhysRevLett.103.016402,PhysRevLett.102.166803}, it exhibits a massless linear dispersion along one principal axis, but massive quadratic dispersion in the other perpendicular axis. Such particular band dispersions make the semi-Dirac materials not only have highly anisotropic transport properties \cite{2016Diffusion}, but also exhibit many interesting physical phenomena, \emph{e.g.}, non-Fermi liquid behavior \cite{PhysRevLett.116.076803}, unusual Landau levels \cite{PhysRevLett.100.236405,PhysRevB.82.035438}, and quantum thermoelectric effect \cite{PhysRevB.100.081403}. In past few years, some real materials have been confirmed to harbor Dirac nodal line \cite{feng2017experimental,gao2018epitaxial,PhysRevB.97.125312,zhang2017two,PhysRevB.98.115164,chen2018prediction} or semi-Dirac states \cite{PhysRevLett.102.166803,kim2015observation,PhysRevB.74.033413}. And what is more,  the coexistence of them have also been proved to be possible \cite{zhang2017dirac}, which is much needed for the device applications with greater flexibility. However, in the only known hypothetical system hr-sB \cite{kim2016electronic}, the Dirac nodal line and semi-Dirac cone are mixed with each other in the same energy window. Naturally, this leaves two pending questions: whether these anisotropic Dirac states can coexist with energy separation around the Fermi level in a real material that has been achieved in experiment, and whether the states can be tuned by applying external feasible means.

Beyond the traditional organometallic frameworks, Chi \emph{et al.} \cite{2022Substrate} recently synthesized an interesting 2D metal-organic hybrid, Ag$_3$C$_2$$_0$ monolayer (ML), which is assembled by organic molecules linked via multiple aryl-metal bonds (\emph{bay}-aryl-metal and \emph{peri}-aryl-metal bonds) rather than single aryl-metal bonds. The incorporation of multiple aryl-metal bonds not only increase the diversity of 2D metal-organic hybrids, but also raise a question worthy of exploring: can the coexistence of multiple aryl-metal bonds bring interesting anisotropic physics, especially in aspect of the above mentioned Dirac electronic states. Inspired by this, using first-principles calculations, we systematically investigate the mechanical and electronic properties of Ag$_3$C$_2$$_0$ ML. Our results show that Ag$_3$C$_2$$_0$ ML is strongly anisotropic in both mechanics and electronics. Remarkably, both of Dirac nodal line and semi-Dirac states can be obtained in Ag$_3$C$_2$$_0$ ML in an easily tunable way, and they reside on the different sides of the vicinity of the Fermi level without the overlap of energy. Therefore, it should be possible to realize the switch between these two types of Dirac states through opposite gate voltages, and achieve the desirable anisotropic carrier transport in future Ag$_3$C$_2$$_0$ ML based nanodevices.

\section{COMPUTATIONAL METHODS}
All our first-principles density functional calculations are performed in the Vienna \textit{Ab} \textit{initio} Simulation Package 
\cite{PhysRevB.54.11169}. The projector augmented wave (PAW) method \cite{PhysRevB.59.1758} is adopted to deal with the core-valence interactions, taking a cutoff energy of 500 eV for the plane-wave basis. The Perdew-Burke-Ernzerhof (PBE) \cite{PhysRevLett.77.3865} functional within the generalized gradient approximation (GGA) is employed to describe the exchange-correlation interactions. The Monkhorst-Pack \textit{\textbf{k}}-point mesh with a uniform spacing of 
$2{\pi}{\times}$0.02 {\AA}$^{-1}$ is used to sample the 2D BZ. To avoid the artificial interactions 
between periodic images, a 20 {\AA} vacuum layer is applied along the perpendicular direction of the 2D atomic structure. The convergence criterion of electronic iteration is set to $10^{-6}$ eV, and the structure is relaxed until Hellmann-Feynman force on each atom is less than 0.001 eV/{\AA}.For the electronic bands calculations, the more accurate hybrid functional HSE06 \cite{heyd2003hybrid} is also employed to validate the PBE results.

\section{RESULTS AND DISCUSSION}
\subsection{Atomic Structure and Bonds}
\begin{figure}[t]
	\includegraphics[width=1.\linewidth]{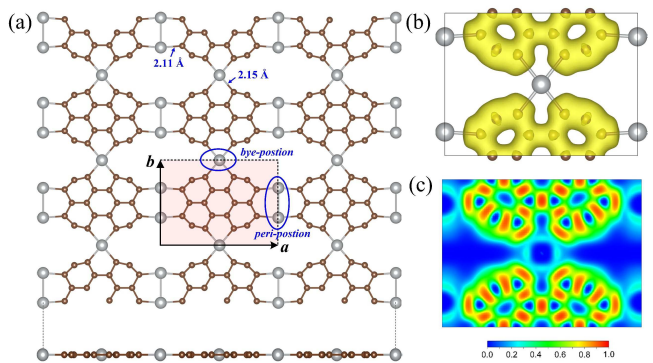}
	\caption{\label{fig:1}(a) Top and side views for the optimized atomic structure of Ag$_3$C$_2$$_0$ ML. The magenta shadow area denotes the corresponding primitive cell. (b) The top view of the electron localization function (ELF) for Ag$_3$C$_2$$_0$ ML with the isosurface of 0.72. (c) The corresponding 2D slice of ELF. Here, the site with ELF = 0 has no electron distribution, while that with ELF = 1.0 holds a completely localized electron.}
\end{figure}

Without any lattice constraint, the freestanding Ag$_3$C$_2$$_0$ ML is fully relaxed to be a completely planar structure in an orthorhombic lattice with \textit{Pmmm} symmetry (No. 47), as shown in Figs. \ref{fig:1}a. The equilibrium lattice parameters are optimized to be $a = 1.14$ nm and $b = 0.83$ nm, which are quite consistent with the experiment results \cite{2022Substrate}, $a =1.16 \pm 0.02$ nm and $b=0.86 \pm 0.02$ nm, respectively. This means that the selected computational strategy is suitable for Ag$_3$C$_2$$_0$ ML well. In the primitive cell (see the shadow area in Fig. \ref{fig:1}a), there is a complete perylene \cite{aizenshtat1973perylene} skeleton formed by five C6 rings with 20 C atoms, sharing four Ag atoms at the \emph{peri}-positions along \emph{\textit{\textbf{a}}}-axis and two Ag atoms at the \emph{bay}-positions along \emph{\textit{\textbf{b}}}-axis. Thus, there exists two kinds of aryl-metal bonds \cite{2022Substrate}, \emph{i.e.}, \emph{peri}-Ag-C bonds (2.11 {\AA}) and \emph{bay}-Ag-C bonds (2.15 {\AA}), interlinking the isolated perylene motifs to constructure the framework of Ag$_3$C$_2$$_0$ ML. The calculated electron localization function (ELF, see Figs. \ref{fig:1}b and \ref{fig:1}c), being able to  reflect the degree of electron localization \cite{silvi1994classification},suggest that both \emph{peri- and bay}-Ag-C bonds belong to the ionic interaction with C atoms as the acceptor. As a whole, the strength of interactions along the \emph{\textit{\textbf{a}}}-axis depend on the ionic \emph{peri}-Ag-C and armchair covalence C-C bonds in perylene motifs, and that along the \emph{\textit{\textbf{a}}}-axis up on the \emph{bay}-Ag-C and zigzag C-C bonds. In view of such interesting structural features, hereafter we’ll focus on the anisotropy of Ag$_3$C$_2$$_0$ ML in both mechanical and electronic properties, and the possible coupling between them.

\subsection{Mechanical Anisotropy}
\begin{figure}[t]
	\includegraphics[width=1.\linewidth]{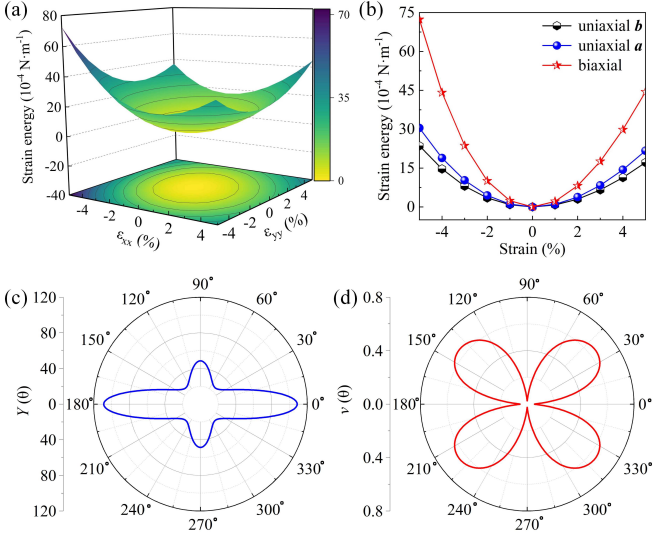} \caption{\label{fig:2} (a) The stain energies $U(\varepsilon)$ per unit area for Ag$_3$C$_2$$_0$ ML with respect to a $\varepsilon_{x x}$-$\varepsilon_{y y}$ stain plane in the range of $-$5\%$\sim$5\%, together with the corresponding 2D contour plot. (b) The stain energies $U(\varepsilon)$ as a function of the uniaxial $\emph{\textit{\textbf{a}}}/\emph{\textit{\textbf{b}}}$ and biaxial strains. As is shown, the easiest energy changing is along the biaxial strains, followed by the uniaxial  \emph{\textit{\textbf{b}}} strains, and then the uniaxial \emph{\textit{\textbf{a}}} strains.} 	
\end{figure}

To explore the mechanical properties of Ag$_3$C$_2$$_0$ ML, we first examine its elastic strain energy $U(\varepsilon)$ per unit area caused by in-plane strains, which can be expressed as, \cite{zhang2015penta}
\begin{equation}\label{func_1}
	U(\varepsilon)=\frac{1}{2} C_{11} \varepsilon_{x x}^{2}+\frac{1}{2} C_{22}
	\varepsilon_{y y}^{2}+\frac{1}{2} C_{12} \varepsilon_{x x} \varepsilon_{y y}+2
	C_{66} \varepsilon_{x y}^{2},
\end{equation}
for a 2D system. Thereinto, the standard Voigt notation \cite{PhysRevB.85.125428} (1-xx, 2-yy, and 6-xy) is used. Thus, the elastic constants \textit{C}$_{ij}$ can be derived by fitting the strain energy curves under different strains $\varepsilon_{ij}$. Based on the obtained $U(\varepsilon)$ of Ag$_3$C$_2$$_0$ ML (Figs. \ref{fig:2}a and \ref{fig:2}b), one see that the energy response to the applied strains is anisotropic, which can also be reflected by the large difference among the fitted elastic constants (\emph{e.g.}, \textit{C}$_{11}$ = 108.83 N/m $\sim$ \textit{C}$_{12}$ = 2.26 N/m, and \textit{C}$_{22}$ = 48.65 N/m $\sim$ \textit{C}$_{66}$ = 7.57 N/m). To demonstrate the directional dependence of the mechanical strength in details, we then evaluate the Young's modulus $\textit{Y}(\theta)$ and Poisson's ratio $\textit{v}(\theta)$ from the obtained elastic constants \cite{PhysRevB.82.235414}:
\begin{equation}\label{func_1}
	\textit{Y}(\theta)=\frac{{C}_{11} {C}_{22}-{C}_{12}^{2}}{{C}_{11} {~s}^{4}+\left(\frac{{C}_{11} {C}_{22}-{C}_{12}^{2}}{{C}_{66}}-2 {C}_{12}\right) {s}^{2} {c}^{2}+{C}_{22} {c}^{4}}, 
\end{equation}
\begin{equation}\label{func_1}
	\textit{v}(\theta)=-\frac{\left({C}_{11}+ {C}_{22}-\frac{ {C}_{11} {C}_{22}- {C}_{12}^{2}}{ {C}_{66}}\right) {s}^{2} {c}^{2}- {C}_{12}\left({~s}^{2}+ {c}^{4}\right)}{{C}_{11} {~s}^{4}+\left(\frac{ {C}_{11} {C}_{22}- {C}_{12}^{2}}{ {C}_{66}}-2 {C}_{12}\right) {s}^{2} {c}^{2}+ {C}_{22} {c}^{4}}
\end{equation}
where ${\theta}$ is the polar angle with respect to \emph{\textit{\textbf{a}}}-axis; \emph{s} and \emph{c} denote {$\sin (\theta)$} and {$\cos (\theta)$}, respectively. The calculated $\textit{Y}(\theta)$ and $\textit{v}(\theta)$ of Ag$_3$C$_2$$_0$ ML are displayed in Figs. \ref{fig:2}c and \ref{fig:2}d, respectively. With the increasing of ${\theta}$, ranging from $0^{\circ}$ (\emph{\textit{\textbf{a}}}-axis) to $90^{\circ}$ (\emph{\textit{\textbf{b}}}-axis), the Young's modulus firstly decreases from the maximal value (108.69 N/m at ${\theta}$ = $0^{\circ}$) to the smallest value (24.86 N/m at ${\theta}$ = $45^{\circ}$), and then increase to a relatively small maximum (48.58 N/m at ${\theta}$ = $90^{\circ}$). The ratio between the maximum and minimum of Young's modulus is about 4.41, obviously larger than that of W-graphane ($\sim$2.62) \cite{PhysRevB.82.235414}, palgraphyne ($\sim$3.29) \cite{zhao2020palgraphyne}. In this sense, Ag$_3$C$_2$$_0$ ML is a new strong mechanically anisotropic 2D material. Moreover, from the Poisson's ratios (Fig. \ref{fig:2}d) one can get the consistent conclusion. In contrast to the Young’s modulus, the Poisson's ratios, appearing as a flower petal, have the maximal values (0.652) in the ${\theta}$ = $45^{\circ}$ direction, while hold the vanishingly small values along the principle \emph{\textit{\textbf{a}}}- and \emph{\textit{\textbf{b}}}-axis.

\subsection{Topological Nodal lines and Direction-Dependent Dirac Dispersions }
\begin{figure}[t]
	\includegraphics[width=1.\linewidth]{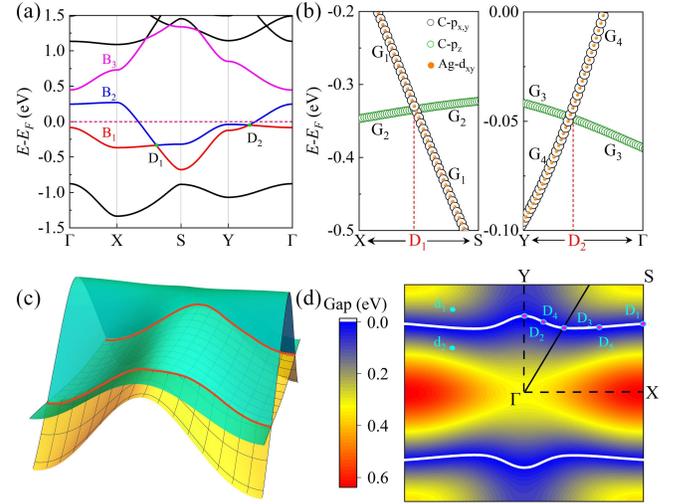} \caption{\label{fig:3} (a) The electronic band structure of intrinsic Ag$_3$C$_2$$_0$ ML. (b) The orbital and atom decomposed bands with irreducible representations around the Dirac points of D$_1$ and D$_2$. (c) The 3D band structure in the entire first Brillouin zone. (d) The contour plot of the energy bandgap between bands B$_1$ and B$_2$.} 	
\end{figure}

The electronic band structure, as depicted in Fig. \ref{fig:3}a, reveals that Ag$_3$C$_2$$_0$ ML is a metal having a partially filled conductive band. Closing to the Fermi level blow, the denoted bands B$_1$ and B$_2$ meet at two points, namely D$_1$ (0, 0.3403) at the path of $\Gamma \rightarrow \mathrm{Y}$ and D$_2$ (0.5, 0.30723) at $\mathrm{X} \rightarrow \mathrm{S}$, with the low-energy bands exhibiting linear dispersion. Since one of the linear bands is almost flat, the band dispersion around both D$_1$ and D$_2$ can be referred to as quasi type-III Dirac dispersion, which appears in a certain direction of type-III Dirac cone \cite{PhysRevLett.120.237403,PhysRevB.98.121110}, and the transverse direction of type-III nodal line \cite{PhysRevB.101.100303}. Note that the two Dirac points and the quasi flat bands can be reproduced in a more accurate HSE06 calculation (see Fig. S1). To understand the origin of the found Dirac states, we calculate the orbital and atom decomposed bands near D$_1$ and D$_2$.  As illustrated in Fig. \ref{fig:3}b, the wave functions near the Dirac points are mainly contributed by \emph{p$_{x,y}$} and \emph{p$_z$} orbitals of C atoms, together with few Ag-\emph{d$_{xy}$} orbitals. Obviously, the energy ordering exchange occurs between \emph{p$_{x,y}$} and \emph{p$_z$}. In fact, band inversion like this is one of the key features for topological semimetals or metals \cite{liu2018all,weng2016topological}.

\begin{figure*}[t]
	\includegraphics[width=0.7\linewidth]{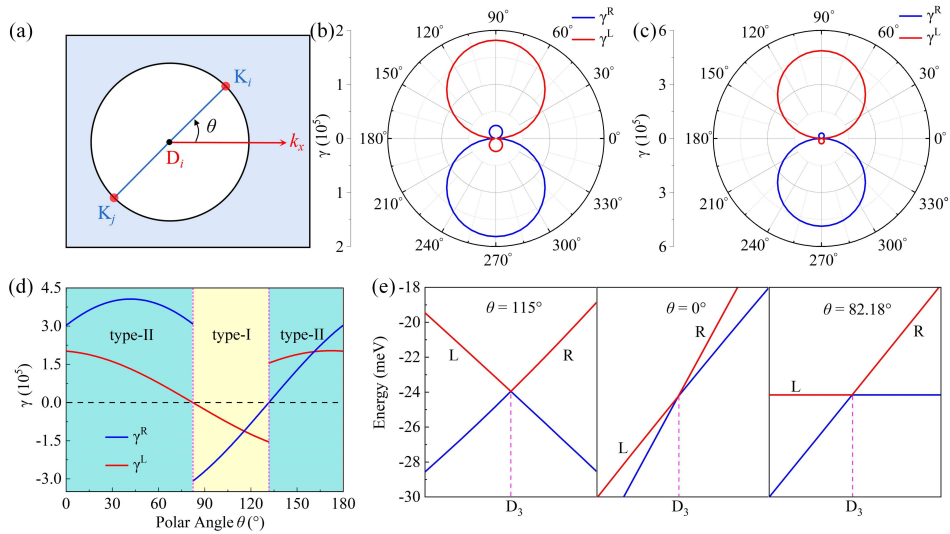}
	\caption{\label{fig:4}(a) The schematic diagram of the polar coordinate in the 2D momentum space with the studied Dirac point \emph{D$_i$} (\emph{i} =1, 2, 3) as the the origin. The distribution of the slope indexes $\gamma^{\mathrm{L}}$ and $\gamma^{\mathrm{R}}$ along different \textit{\textbf{k}}-directions (${\theta}$) for (b) D$_1$, (c) D$_2$ and (d) D$_3$. (e) The low-energy band structures around the Dirac point D$_3$ along the \textit{\textbf{k}}-paths of ${\theta}$ = $115^{\circ}$, $0^{\circ}$ and $82.18^{\circ}$. }
\end{figure*}

The identified multi-crossing Dirac points are reminiscent of the Dirac nodal lines \cite{2017Line,2017dirac}, which is composed of consecutive Dirac points. To reveal the complete pattern of the Dirac points, we plot the 3D band structure of Ag$_3$C$_2$$_0$ ML in the entire BZ, as presented in Fig. \ref{fig:3}c. The result confirms that B$_1$ and B$_2$ bands meet with each other along two open antiparallel nodal lines, running across the whole BZ (Fig. \ref{fig:3}d) along the \emph{\textit{\textbf{a}}}-axis direction. Since all the band dispersions along the transverse direction of the nodal lines are belong to quasi type-III, one can conclude that Ag$_3$C$_2$$_0$ ML is a quasi type-III nodal line metal. In order to study how these nodal lines form, we further carry out the symmetry analysis for the crossing B$_1$ and B$_2$ bands of Ag$_3$C$_2$$_0$ ML using the IRVSP code \cite{gao2021irvsp}. Around D$_1$ and D$_2$ in the highly symmetric paths, the obtained irreducible representations of the two crossing bands are G$_1$ and G$_2$, and G$_3$ and G$_4$, respectively (see Fig. \ref{fig:3}b). As for other general Dirac points (\emph{e.g.}, the arbitrary D$_3$ and D$_4$ in Fig. \ref{fig:3}d), which are not on the high symmetry line, the identified representations are all G$_1$ and G$_2$, as shown in Fig. S2. With the help of Bilbao Crystallographic Server \cite{elcoro2017double}, we confirm that common symmetrical operator corresponding to these representations is \emph{m$_{010}$} mirror symmetry, belonging to the second-order point group \textit{C}$_{s}$. Therefore, the nodal lines in Ag$_3$C$_2$$_0$ ML are caused by the band inversions between C-\emph{p$_{x,y}$} and C-\emph{p$_z$}, protected by mirror reflection symmetry.

Owing to the mirror symmetry protections mechanism, the topological properties of Ag$_3$C$_2$$_0$ ML should be charactered by the topological invariant $\zeta$$_0$, which is suitable for type-A topological nodal line semimetals \cite{Topological2016Fang}. In the BZ plane, we first pick up two \textit{\textbf{k}}-points (\emph{i.e.}, d$_1$ and d$_2$, see Fig. \ref{fig:3}d) on different sides of the nodal line. Subsequently, we count the number of bands below the energy of Dirac point with the mirror eigenvalue of $+1$ at d$_1$ and d$_2$ (see Table S1), which are denoted by N$_1$ (41) and N$_2$ (40), respectively.  Finally, the topological invariant is evaluated to be 1 using $\zeta$$_0$ = N$_1$ $-$ N$_2$. Hence, the linear crossings of the nodal lines in Ag$_3$C$_2$$_0$ ML should not be accidental——as long as the mirror symmetry is preserved, they will note be opened up by an arbitrarily small perturbation.  

To demonstrate the anisotropy of the lower-energy band dispersions around nodal points in the nodal line (taking the D$_1$, D$_2$ and D$_3$ as examples, see Fig. \ref{fig:3}d), we further calculate the band structures along the possible \textit{\textbf{k}} paths except for the tangent directions of the nodal line. As schematized in Fig. \ref{fig:4}a, 1000 \textit{\textbf{k}}-paths, characterized by K\emph{$_i$} → K\emph{$_j$} and polar angle ${\theta}$, are sampled uniformly in a polar coordinate system with respect to the \emph{k$_x$}-axis. On the basis of the obtained 1000 band structures, the slope index is employed to describe the band dispersion features, which is defined as $\gamma^{i} = \emph{ v}{_{F}}^\emph {i} \operatorname{sgn}\left(\emph{v}{_{F}}^\emph {i}\right)$ \cite{PhysRevB.102.155133}. Here \emph {i} (= L or R) indicates the two Dirac bands, including left (L) and ring (R) bands, like that in Fig. \ref{fig:4}e; $\emph{ v}{_{F}}^\emph {i}$ represents the Fermi velocity for the corresponding linear bands, which can be evaluated from the equation of $\hbar v_{\mathrm{F}}=d E(\boldsymbol{k}) / {d} \boldsymbol{k}$. Figs. \ref{fig:4}b and \ref{fig:4}c show the slope indexes $\gamma^{i}$ as a function of the polar angle ${\theta}$ for the low-energy bands around D$_1$ and D$_2$, respectively. Two common characteristics can be seen: (i) both $\gamma^{\mathrm{R}}$ and $\gamma^{\mathrm{L}}$ are positive in all of the considered \textit{\textbf{k}}-paths, which means that the slopes of the two bands share the same sign, and the Dirac dispersion should belong to type-I (or quasi type-III, see following characteristic) along the corresponding \textit{\textbf{k}}-path; (ii) the dispersion of the two bands is quite different——one of them is almost dispersionless, being far less than the other. As for the case of D$_3$, the most important feature is that there exist a change in the sign of $\gamma^{i}$, as illustrated in Fig. \ref{fig:4}d. Namely, the type of Dirac dispersions have been completely changed with the change of the \textit{\textbf{k}}-paths. Specifically, as ${\theta}$ increases  from $0^{\circ}$ to $180^{\circ}$, the band dispersion change from type-I to type-II, and then to type-I again. In the transition boundary between them, the type-III dispersion with one dispersionless band can be found at two critical directions ($\sim$$82.18^{\circ}$ and $\sim$$131.9^{\circ}$). The more direct visualization can be seen in the low-energy band structures along three representative directions (see Fig. \ref{fig:4}e), \emph{i.e.}, ${\theta}$ = $0^{\circ}$, $82.18^{\circ}$ and $115^{\circ}$, which exhibit the features of type-I, type-II and type-III Dirac dispersions, respectively.

\begin{figure*}[t]
	\includegraphics[width=0.7\linewidth]{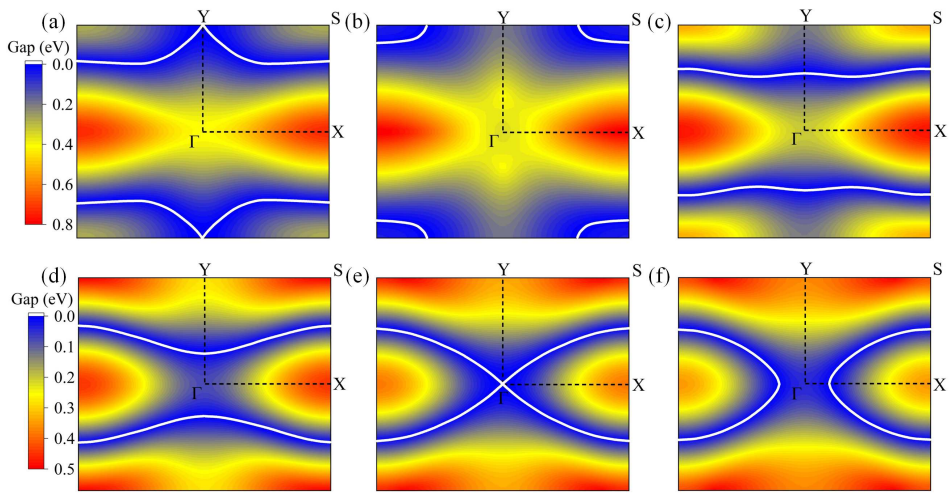}
	\caption{\label{fig:5}The contour plot of the energy bandgap between bands B$_1$ and B$_2$. (a) under $-$0.8\% biaxial strain. (b) under $-$3\% biaxial strain. (c) under 1\% uniaxial \emph{\textit{\textbf{b}}} strain. (d) under 3\% uniaxial \emph{\textit{\textbf{b}}} strain. (e) under 4.2\% uniaxial \emph{\textit{\textbf{b}}} strain. (f) under 5\ uniaxial \emph{\textit{\textbf{b}}} strain. }
\end{figure*}

\subsection{Strains Induced Nodal Line Evolution and Semi-Dirac State.}
As is known, applying mechanical strain is an effective approach for engineering the electronic properties of 2D materials. \cite{deng2018strain,si2016strain} In consideration of the strong anisotropy of mechanical properties, we apply uniaxial (\emph{\textit{\textbf{a}}}-axis or \emph{\textit{\textbf{b}}}-axis) and biaxial strains ranging from $-$5\%$\sim$5\% (a modest range in experiments \cite{deng2018strain}) for Ag$_3$C$_2$$_0$ ML, and  then relax all the atomic positions by fixing the corresponding lattice. On the basis of the optimized new structures, we recalculate their electronic band structures, as illustrated in Figs. S3-S5. The results reveal that the electronic bands of Ag$_3$C$_2$$_0$ ML is mechanically sensitive. Under different strains, some intriguing electronic states emerge around the Fermi level.  In the following, we will focus on two aspects: (i) the evolution of nodal lines formed by B$_1$ and B$_2$ blow the Fermi level, and (ii) the Dirac cone and Semi-Dirac state formed by B$_2$ and B$_3$ above the Fermi level.

\emph{The evolution of the nodal lines.} Under the three types compressive strains, the two nodal lines all shift along opposite direction along \emph{k$_y$} axis. After they touch the boundary of BZ at the high symmetry  Y or Y' point under a critical strain (about 0.8\%, 1.5\%, and 2.2\% for biaxial, uniaxial \emph{\textit{\textbf{b}}} and uniaxial \emph{\textit{\textbf{a}}} strains, respectively), their geometry transform from the two open lines into one closed loop (see Fig. \ref{fig:5}a,b). More specifically, it is a hybrid nodal loop centered at the corner point S, exhibiting both type-III and type-II Dirac dispersions along its transverse directions, as is displayed in the bands along $\mathrm{X} \rightarrow \mathrm{S}$ and $\mathrm{Y} \rightarrow \mathrm{S}$, respectively (Figs. S3-S5).  With the increase of compressive strains, the nodal loop will then gradually shrink. By biaxial strains, the loop will eventually disappear when the strain is larger than 3.5\%, which, however, will not happen in the uniaxial cases due to the anisotropy.

As for uniaxial a tensile strains, the quasi-type-III nodal lines are almost unchanged. However, the interesting evolutions can be found in the other two types of strains. With the increase of the uniaxial \emph{\textit{\textbf{b}}} and biaxial strains, the two nodal points in the \emph{k$_y$} axis gradually moves close to $\Gamma$, and meet with each other at the center of BZ under $\sim$4.2\% and $\sim$5\% for uniaxial \emph{\textit{\textbf{b}}} and biaxial strains, respectively (Fig. \ref{fig:5}e). On the contrast  the nodal points in $\mathrm{X} \rightarrow \mathrm{S}$ and $\mathrm{X'} \rightarrow \mathrm{S'}$ are almost fixed. Because of such difference in mechanical response, the nodal lines can be tuned from two antiparallel curves protruding towards Y, to two completely straight lines (Fig. \ref{fig:5}c), and then to two curves protruding towards the opposite direction (Fig. \ref{fig:5}d). Moreover, under uniaxial \emph{\textit{\textbf{b}}} strains ($> 4.2\%$) (Fig. \ref{fig:5}f), the two open lines can also evolve into one closed loop, but centered at the $\mathrm{X}/\mathrm{X'}$ points. 

\begin{figure}[t]
	\includegraphics[width=1.\linewidth]{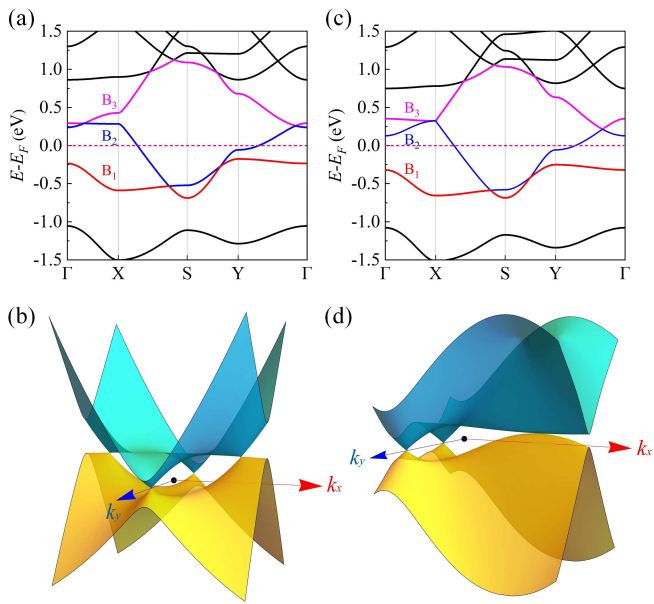}
	\caption{\label{fig:6}(a) The electronic band structure of Ag$_3$C$_2$$_0$ ML under $-$2\% biaxial strain. (c) The electronic band structure of Ag$_3$C$_2$$_0$ ML under $-$3\% biaxial strain. (b) and (d) The 3D plots of B$_2$ and B$_3$ bands. }
\end{figure}

\emph{Dirac cones and semi-Dirac state.} Finally, let's turn our attention to B$_2$ and B$_3$ bands near the Fermi level above. There is a sizable bandgap between B$_2$ and B$_3$ bands for all the considered tensile and compressive uniaxial \emph{\textit{\textbf{a}}} strains. Under modest uniaxial \emph{\textit{\textbf{b}}}  and biaxial strains, however, the two gapped bands will crossing with other, and form some intriguing isolated Dirac cones. Specifically, two pair of isolated Dirac cones (Figs. \ref{fig:6}a and \ref{fig:6}b) can be induced when the compressive biaxial (uniaxial \emph{\textit{\textbf{b}}}) strain is larger than $\sim$1.2\% ($\sim$2.0\%). As the strain continues to increase, the two coupled Dirac cones along \emph{k$_x$} axis merge together and form a semi-Dirac cone \cite{PhysRevLett.103.016402,PhysRevLett.102.166803} at the X point (see Figs. \ref{fig:6}c and \ref{fig:6}d) under a critical strain of 3\% for the biaxial case (4.2\% for uniaxial \emph{\textit{\textbf{b}}}). Unlike the conventional Dirac cone, the semi-Dirac state exhibits the strong dispersion anisotropy, having the standard type-I Dirac dispersion in the direction of $\mathrm{X} \rightarrow \mathrm{S}$, but quadratic dispersion in the perpendicular $\mathrm{X} \rightarrow \Gamma$ path. What makes it unique is that one of the “quadratic” bands is almost dispersionless and linear, shaped like a flat band. In this regard, the identified semi-Dirac cone of the strained Ag$_3$C$_2$$_0$ ML should be referred to as \emph{double semi-Dirac cone}. In practical application, three types of fermions with quite different carrier transport velocities should be obtainable in their respective directions, including superheavy localized fermions, conventional massive carries, and ultrafast massless Dirac fermions. With the strains increasing further, the semi-Dirac cone disappear, but the two isolated Dirac cone along the \emph{k$_y$} axis can still be preserved up to 5\% compressive strains (both uniaxial \emph{\textit{\textbf{b}}}) and biaxial strains)
\section{CONCLUSION}
In summary, based on the recently synthesized Ag$_3$C$_2$$_0$ ML, we perform a systematical first-principles study about its mechanical and electronic properties. Our results show that Ag$_3$C$_2$$_0$ ML has strong mechanical anisotropy due to its particular structure constructed by multiple aryl-metal bonds. Associated with this, it also holds two types of desirable Dirac states, including Dirac nodal line and semi-Dirac cone with highly anisotropic band dispersions. Specifically, its intrinsic Dirac points form two antiparallel nodal lines near the Fermi level blow. Moreover, around the certain Dirac points of the nodal lines, type-I, -II and -III Dirac dispersions can all be found in the low-energy bands. Under a modest strain, the open nodal lines can evolve into a closed hybrid nodal loop, and a special semi-Dirac cone can be induced above the Fermi level, which holds multiple direction-dependent fermions, including superheavy localized fermions, normal massive carries, and ultrafast massless Dirac fermions. Given these, it is reasonable to believe that the synthesized Ag$_3$C$_2$$_0$ ML should be a promising platform for constructing the much flexible nanodevices exhibiting interesting topological physics and highly anisotropic Dirac carries transport.

\section*{ACKNOWLEDGMENTS}
This work is supported by the National Natural Science Foundation of China 
(11964023), Natural Science Foundation of Inner Mongolia Autonomous Region 
(2021JQ-001), and the 2020 Institutional Support Program for Youth Science and 
Technology Talents in Inner Mongolia Autonomous Region (NJYT-20-B02). 
~\\
\bibliography{Ag3C20-manuscript}

\end{document}